\documentclass[aps,12pt, a4paper, preprint]{revtex4-1}

\usepackage[normalem]{ulem}
\usepackage[utf8]{inputenc}
\usepackage[T1]{fontenc}
\usepackage{amsmath}
\usepackage{amsfonts}
\usepackage{amssymb}
\usepackage{graphics,graphicx}

\usepackage{graphicx,color}
\usepackage{subfigure} 
\usepackage{wrapfig} 
\usepackage{epstopdf}
\graphicspath{{figuras/}}

\usepackage[breaklinks=true]{hyperref}
\usepackage{setspace}

\newcommand{\bes}{\begin{subequations}}
\newcommand{\ees}{\end{subequations}}
\def\ben{\begin{eqnarray}}
\def\een{\end{eqnarray}}
\def\be{\begin{equation}}
\def\ee{\end{equation}}

\begin{document}
 
\title{Black branes in asymptotically Lifshitz spacetimes \\ in $\kappa$-deformed Horndeski gravity}

\author{Fabiano F. Santos$^{1}$}
\email[Eletronic address:]{fabiano.ffs23@gmail.com}
\author{Henrique Boschi-Filho$^{2}$}
\email[Eletronic address:]{boschi@if.ufrj.br}  
\affiliation{  
$^1$Departamento de Física, Universidade Federal do Maranhão, \\ Campus Universitario do Bacanga, São Luís (MA), 65080-805, Brazil\\
$^2$Instituto de F\'{\i}sica, Universidade Federal do Rio de Janeiro, \\ 21.941-909, Rio de Janeiro - RJ, Brazil}

\begin{abstract}
In this work we consider Horndesky gravity deformed by a parameter in its kinetic terms embedded in asymptotically Lifshitz spacetimes. We obtain black brane solutions in this geometry and study their thermodynamical properties. We show that the $\kappa$-deformation on the Horndeski kinetic terms allow for arbitrary values of the critical exponents characteristic of Lifshitz spacetimes. These solutions present local and global thermodynamical stabilites. 
\end{abstract}

\maketitle

\tableofcontents

\section{Introduction}

Horndeski gravity  is a scalar-tensor theory with second-order field equations and second-order energy moment tensor \cite{Horndeski:1974wa, Charmousis:2011bf, Charmousis:2011ea, Bruneton:2012zk, Heisenberg:2018vsk, Kobayashi:2019hrl}. 
Since its discovery in 1974, it has aroused great interest in its applications ranging from experimental and theoretical cosmology to more formal aspects of general relativity. 
 The recent detection by LIGO \cite{LIGOScientific:2017ync} of the two neutron stars merging event GW170817 has brought new speculations on the relevance of the {\sl John} sector of Horndeski gravity \cite{Oikonomou:2020sij}. This sector belongs to a more general class called the {\sl fab four} theory, where  the associated Lagrangian includes four arbitrary functions of the scalar field \cite{Charmousis:2011bf, Charmousis:2011ea}.  
   Different features of Horndeski gravity were discussed, for instance, in \cite{Rinaldi:2012vy,  Rinaldi:2016oqp, Cisterna:2016vdx, Gleyzes:2013ooa, Zumalacarregui:2013pma, Starobinsky:2016kua, Brito:2018pwe}.

A landmark of present days theoretical physics is the AdS/CFT correspondence \cite{Maldacena:1997re, Aharony:1999ti, Petersen:1999zh, Ramallo:2013bua} which relates a gravity theory in $D+1$ dimensions with quantum field theories in $D$ dimensions. This duality started with string/M theory and have found applications in a wide range of physical set ups from the quark-gluon plasma (QGP) \cite{Policastro:2001yc, Kovtun:2003wp, Kovtun:2004de} to condensed matter systems (CMS) \cite{Hartnoll:2009sz, Sachdev:2011wg}. So, considering extensions of the AdS/CFT correspondence, one can related Horndeski gravity in asymptotic AdS spacetimes with many interesting properties as, for instance, generalized results for black hole and black brane solutions and thermodynamics \cite{Anabalon:2013oea, Cisterna:2014nua, Cisterna:2017jmv}, 
and the viscosity bound of the dual QGP \cite{Feng:2015oea}.  

The importance of Lifshitz spacetimes to classical and quantum gravity was put forward in the seminal works by Ho\v{r}ava \cite{Horava:2008ih, Horava:2009if, Horava:2010zj}. A general characteristic of this theory is that it has critical  exponents $z$. For $z\not=1$ the Lorentz symmetry is violated, leading to applications from black holes \cite{Barausse:2011pu, Wang:2017brl} and cosmology \cite{Clifton:2011jh} to condensed matter \cite{Hartnoll:2009sz, Sachdev:2011wg}. In the particular case of $z=1$ it is directly connected with the AdS/CFT correspondence in Poincaré coordinates \cite{Maldacena:1997re, Aharony:1999ti, Petersen:1999zh, Ramallo:2013bua}.

Asymptotically AdS  black holes in Lifshitz spacetimes were discussed, for instance, in \cite{Danielsson:2009gi, Bertoldi:2009vn} and extended in \cite{Bravo-Gaete:2013dca, Santos:2020egn,  Brito:2019ose} to include Horndeski gravity. In particular, in Refs. \cite{Santos:2020egn,  Brito:2019ose} it were  considered the case of black branes in Horndeski gravity finding solutions with particular values of the critical exponents.  

Here, we reanalyse black brane thermodynamics in asymptotically AdS Lifshitz spacetimes within Horndeski gravity modified by the recently proposed $\kappa$-deformation \cite{Santos:2022fbq} of the corresponding kinetic terms. We show that this proposal leads to a generalized entropy for black branes  compatible with  the area law. The introduction of the $\kappa$-deformation releases the constraints of Horndeski theory allowing for arbitrary values of the critical exponents. 

This work is summarized as follows: In Sec.~\ref{v1}, we briefly review Horndeski theory and introduce the covariant $\kappa$-deformed Horndeski theory. In Sec.~\ref{v2}, we present the ansatz for black brane solutions in  asymptotic Lifshitz spacetimes in four dimensions with arbitrary critical exponents. The  thermodynamics of these solutions are  we  analyzed in Sec.~\ref{v3}, varying the critical exponents of the relevant quantities, and finally, in Sec.~\ref{v4}, we present our conclusions.


\section{Covariant $\kappa$-deformed Horndeski Theory}\label{v1}

Let us start this section briefly reviewing the Horndeski gravity defined by the following Lagrangian 
\begin{eqnarray}\label{LH}
{\cal L}_{\rm Horndeski} = {\cal L}_{\rm EH} + \sum_{n=2}^{5}{\cal L}_n \,, 
\end{eqnarray}
\noindent where ${\cal L}_{\rm EH}$ is the Einstein-Hilbert Lagrangian, while  \cite{Charmousis:2011bf}
\begin{eqnarray}
{\cal L}_2 &=& G_2(X, \phi)\,, \nonumber \\ 
{\cal L}_3 &=& -G_3(X, \phi) \Box \phi\,, \nonumber \\ 
{\cal L}_4 &=& -G_4(X, \phi) R + \partial_X G_4(X, \phi) \delta^{\mu \nu}_{\alpha \beta} \nabla^{\alpha}_{\mu} \phi \nabla^{\beta}_{\nu} \phi\,, \nonumber \\ 
{\cal L}_5 &=& -G_5(X, \phi) G_{\mu \nu} \nabla^{\mu} \nabla^{\nu}\phi - \frac{1}{6} \partial_X G_5(X, \phi) \delta^{\mu \nu \rho}_{\alpha \beta \gamma} \nabla^{\alpha}_{\mu} \phi \nabla^{\beta}_{\nu} \phi \nabla^{\gamma}_{\rho} \phi \,. \label{4L}
\end{eqnarray}
The functions $G_n$ of the scalar field $\phi$ and $X$ are arbitrary, while $X$ is defined in terms of the covariant derivatives of $\phi$ as $X \equiv - \frac{1}{2} \nabla_{\mu} \phi \nabla^{\mu} \phi$, and  $G_{\mu \nu}$ is the usual Einstein tensor. 

Now, let us consider the introduction of a covariant $\kappa$-deformation in the Horndeski theory. 

The non-relativistic $\kappa$-deformation presented by \cite{daCosta:2020mbf,Kaniadakis} is derived through a kinetic interaction principle.  In that context, the $\kappa$-derivative is defined in flat space by \cite{Kaniadakis} 
\begin{eqnarray}
\label{eq:q-derivative-dual}
\begin{array}{lll}
  \displaystyle D_{\kappa,x} f(x) 
        &\equiv &      \displaystyle\sqrt{1+\kappa^{2}x^{2}}\;  \frac{df(x)}{dx}.
 \end{array}
\end{eqnarray}

Here, we consider a generalization of the above flat space derivative to a curved spacetime  $\kappa$-deformation of the relativistic covariant derivative \cite{Santos:2022fbq} 
 \begin{equation}\label{kappa}
    \nabla^{\nu}_{\kappa}
 \equiv \sqrt{\Theta_{\kappa}(x)}\nabla^{\nu}  \,, 
 \end{equation} 
where $\Theta_{\kappa}(x)$ is a dimensionless scalar function of the spacetime coordinates $x=(x^0,x^1,x^2,x^3)$ to be determined by the solution of Einstein equations. In this sense, applying this covariant $\kappa$-derivative to Horndeski theory, the Lagrangians ${\cal L}_n$, Eqs. \eqref{4L}, are modified accordingly.


Specifically, in this work, we will restrict our interest to a particular case of the Horndeski theory known as the John sector: 
\begin{eqnarray}\label{John}
   {\cal L}_{\rm John} \equiv  {\cal L}_{\rm EH} + {\cal L}_2 &=&  R-2\Lambda -\frac{1}{2}(\alpha g_{\mu\nu}-\gamma\,  G_{\mu\nu})\nabla^{\mu}\phi\nabla^{\nu}\phi\,, 
\end{eqnarray}
where the parameters $\alpha$ and $\gamma$ control the strength of the kinetic couplings. Then, the action of this sector with the covariant $\kappa$-derivate reads 
\begin{eqnarray}
I[g_{\mu\nu},\phi]&=&
\frac{1}{16\pi G}
\int{d^{4}x\sqrt{-g}\; \left[R-2\Lambda-\frac{1}{2}(\alpha g_{\mu\nu}-\gamma G_{\mu\nu})\nabla^{\mu}_{\kappa}\phi\nabla^{\nu}_{\kappa}\phi\right]}\,.\label{2.1}
\end{eqnarray}
Note that the character of the term $\nabla^{\mu}_{\kappa}\phi\nabla^{\nu}_{\kappa}\phi$ is a second-order covariant derivative of the scalar field. Furthermore, this action is invariant under the shift $\phi\to\phi+constant$ and the discrete $\phi\to-\phi$ symmetries. The shift symmetry is a characteristic of Galileons \cite{Deffayet:2011gz} and of the Hornsdeky scalar field \cite{Kobayashi:2019hrl}, which is up held here in presence of the $\kappa$-deformation \eqref{kappa}.

Thus, the $\kappa$-Horndeski-Einstein field equations can be written formally by varying the action (\ref{2.1}), providing
\begin{equation}
G_{\mu\nu}+\Lambda g_{\mu\nu}=\frac{\Theta_{\kappa}(x)}{2k}T_{\mu\nu},\label{4}
\end{equation}
where $k=1/(16\pi\,G)$. Note that $\kappa$-deformation, Eq. \eqref{kappa}, modifies the usual Einstein equations of general relativity by multiplying the energy-momentumm tensor $T_{\mu\nu}$ by the scalar function $\Theta_{\kappa}(x)$. This can be interpreted as an effective spacetime dependent Newton's constant $G_{\rm eff}=G \, \Theta_{\kappa}(x)$. It is important to remark that a similar feature also appears in formulations of quantum gravity (see {\sl e. g.} \cite{Borissova:2022mgd} and references therein for a recent account). 

The specific form of the energy-momentum tensor derived from Eq. \eqref{2.1} can be written as 
\begin{eqnarray}
T_{\mu\nu}=\alpha T^{(1)}_{\mu\nu}+\gamma T^{(2)}_{\mu\nu}\,, 
\end{eqnarray}
where 
\begin{eqnarray}
T^{(1)}_{\mu\nu}&=&\nabla_{\mu}\phi\nabla_{\nu}\phi-\frac{1}{2}g_{\mu\nu}\nabla_{\lambda}\phi\nabla^{\lambda}\phi \,,
\end{eqnarray}
and 
\begin{eqnarray}
T^{(2)}_{\mu\nu}&=&\frac{1}{2}\nabla_{\mu}\phi\nabla_{\nu}\phi R-2\nabla_{\lambda}\phi\nabla_{(\mu}\phi R^{\lambda}_{\nu)}-\nabla^{\lambda}\phi\nabla^{\rho}\phi R_{\mu\lambda\nu\rho}\nonumber\\
&&-(\nabla_{\mu}\nabla^{\lambda}\phi)(\nabla_{\nu}\nabla_{\lambda}\phi)+(\nabla_{\mu}\nabla_{\nu}\phi)\Box\phi+\frac{1}{2}G_{\mu\nu}(\nabla\phi)^{2}\nonumber\\
&&-g_{\mu\nu}\left[-\frac{1}{2}(\nabla^{\lambda}\nabla^{\rho}\phi)(\nabla_{\lambda}\nabla_{\rho}\phi)+\frac{1}{2}(\Box\phi)^{2}-(\nabla_{\lambda}\phi\nabla_{\rho}\phi)R^{\lambda\rho}\right]\,. 
\end{eqnarray}

Then, the $\kappa$-Horndeski-Einstein field equations, Eq. \eqref{4}, imply that the corresponding scalar field equation is given by
\begin{equation}
\Theta_{\kappa}(x)\nabla_{\mu}[(\alpha g^{\mu\nu}-\gamma G^{\mu\nu})\Theta_{\kappa}(x)\nabla_{\nu}\phi]=0.\label{6}
\end{equation}

In the next section, we are going to obtain black brane solutions for the $\kappa$-Horndeski-Einstein equations with a scalar field governed by the above relation and find the corresponding $\Theta_{\kappa}(x)$ function allowed by these constraints.


\section{Black brane solutions in asymptotically Lifshitz spacetimes}\label{v2}

We start this section considering the following {\sl ansatz} for a four-dimensional black brane solution with static spherical symmetry in asymptotically Lifshitz spacetimes
\begin{equation}
ds^{2}=L^{2}\left(-r^{2z}f(r)dt^{2}+r^{2}(dx^{2}+dy^{2})+\frac{dr^{2}}{r^{2}f(r)}\right),\label{7}
\end{equation}
where $z$ is the Lifshitz exponent. If $z=1$ one recovers the usual black brane solutions. These solutions with $z\not=1$ are specially interesting when considering applications to holographic condensed matter systems (see {\sl e. g.} \cite{Hartnoll:2009sz, Sachdev:2011wg}). 

The spherically symmetrical static configurations of certain Galileons with shift-invariance admit a no-hair theorem \cite{Hui:2012qt}. In addition, the no-hair theorem for Galileons requires that the square of the conserved current $J_{\mu}=(\alpha g_{\mu\nu}-\gamma G_{\mu\nu})\nabla^{\nu}_{\kappa}\phi$, defined in (\ref{6}), should not diverge at the horizon. However, one can escape the no-hair theorem, imposing that the conserved current radial component disappears in an identical way without restricting the radial dependence of the scalar field \cite{Bravo-Gaete:2013dca,Rinaldi:2012vy}:
\begin{equation}
\alpha g_{rr}-\gamma G_{rr}=0\label{7.1}.
\end{equation}
  Defining a new field $\psi\equiv \phi^{'}=d\phi/dr$, and following the procedures of \cite{ Brito:2019ose,Santos:2021orr}, it can be shown that this equation together with (\ref{6}) and the {\sl ansatz} \eqref{7} have a solution which can be written as: 
\begin{eqnarray}
{f}(r)&=&1-\left(\frac{r_{+}}{r}\right)^{2z+1},\label{il2}\\
\psi^{2}(r)&=&-\frac{2(1+2z)\xi}{r^{2}f(r)\Theta_{\kappa}(r)},
\,\,\, \xi=\frac{(\alpha+\gamma\Lambda)}{\alpha^2}.\label{3w2}\\
\Theta_{\kappa}(r)&=&1+\kappa^{2}\left(\frac{r}{r_{+}}\right)^{\lambda};\quad\lambda=(z-1)\frac{\alpha-\gamma\Lambda}{\alpha+\gamma\Lambda}.\label{9.1}
\end{eqnarray}
In our case, the inclusion of $\kappa$-deformation produces a solution that does not necessarily impose a specific value of the critical exponent, and the $\kappa$-Horndeski-Einstein field equations (\ref{4}) and (\ref{6}) are satisfied by the equations (\ref{il2})-(\ref{9.1}) for any value of $z$. This is in contrast with 
the corresponding black brane solutions in asymptotically Lifshitz spacetimes in Horndeski gravity without $\kappa$-deformation \cite{Brito:2019ose,Bravo-Gaete:2013dca}. 

Furthermore, the solutions can be asymptotically dS or AdS for $\alpha/\gamma<0$ or $\alpha/\gamma>0$, respectively \cite{Anabalon:2013oea}. Thus, the stability requires that $(\alpha+\gamma\Lambda)$ is not negative, this leads to an interval of the form $-\infty<\gamma\leq\alpha/(-\Lambda)$, which means $\xi\ge 0$. The value of $\xi=0$ corresponds to the condition $\alpha=-\gamma\Lambda$, which is a degenerate theory \cite{Jiang:2017imk} not considered here.  

We have that for $f\to 1$ with $r\to \infty$ it is an asymptotically AdS spacetime. Thus, the further we advance in spacetime, entering the ultraviolet (UV) region for large values of $r$. However, in $r=r_+$, we have both $g_{tt}(r)$ and $\psi^{2}_{\kappa}(r)$, the latter in action (\ref{2.1}) ensures that this is a genuine vacuum solution with $\Theta_{\kappa}(r)=$constant. Furthermore, we have that the surface located at $ r=r_+$ is infinitely shifted to ultraviolet about an asymptotic observer.

So, we have that the temperature of the black hole is given by \cite{Bekenstein:1973ur, Hawking:1975vcx}
\begin{eqnarray}
T&=&\frac{-g'_{tt}(r\rightarrow r_{0})}{4\pi(\sqrt{-g_{tt}(r\rightarrow r_{0})g_{rr}(r\rightarrow r_{0})})}\,,\\
&=&\frac{(1+2z) r^{z}_{+}}{4\pi}.
\label{T}
\end{eqnarray}
In fact, we have that the appearance of a black hole, which has a flat horizon $\Re^{2}$, leads us to a physics in IR that corresponds to placing the invariant theory of scale at a finite temperature \cite{Hartnoll:2009sz}. From now on, we are going to call the solutions \eqref{7} as black branes or black holes interchangeably. 


\section{Thermodymanics}\label{v3}

Now, we present that free energy can be obtained by evaluating Euclidean action as discussed by \cite{Hartnoll:2009sz,Santos:2021orr}. In fact, as already well established, the path integral must include a Gibbons-Hawking boundary term \cite{Gibbons:1976ue} to provide the correct variational problem to the Horndeski gravity \cite{Santos:2021orr}. Thus, for the action (\ref{2.1}), we have euclidean action:
\begin{eqnarray}
\label{EA}
&&I_{E}=-\frac{1}{16\pi G}\int{\sqrt{g}d^{4}x\mathcal{L}}-\frac{1}{8\pi G}\int_{r\to\infty}{d^{3}x\sqrt{\bar{\gamma}}\mathcal{L}_{b}}+\frac{1}{8\pi G}\int_{r\to\infty}{d^{3}x\sqrt{\bar{\gamma}}\mathcal{L}_{ct}},\\
&&\mathcal{L}=(R-2\Lambda)+\frac{\gamma}{2}G_{\mu\nu}\Theta_{\kappa}^\lambda (r) \nabla^{\mu}\phi\nabla^{\nu}\phi\\
&&\mathcal{L}_{GB}=K^{({\bar{\gamma}})}-\Sigma^{(\bar{\gamma})}+\frac{\gamma}{4}\Theta_{\kappa}^\lambda(r)(\nabla_{\mu}\phi\nabla_{\nu}\phi n^{\mu}n^{\nu}-(\nabla\phi)^{2})K^{(\bar{\gamma})}+\frac{\gamma}{4}\Theta_{\kappa}^\lambda (r) \nabla^{\mu}\phi\nabla^{\nu}\phi K^{(\bar{\gamma})}_{\mu\nu}\\
&&{\cal L}_{ct}=c_{0}+c_{1}R+c_{2}R^{ij}R_{ij}+c_{3}R^{2}+b_{1}\Theta_{\kappa}^\lambda (r) (\partial_{i}\phi\partial^{i}\phi)^{2}. 
\end{eqnarray}
Here ${\cal L}_{ct}$ are boundary counterterms, they do not affect the bulk dynamics and will be neglected. $\mathcal{L}_{GB}$ corresponds to the Gibbons-Hawking $\gamma$-dependent terms associated with the Horndeski gravity and $n^{\mu}$ is an outward pointing unit normal vector to the boundary and $K^{(\bar{\gamma})}=\bar{\gamma}^{\mu\nu}K^{({\bar{\gamma}})}_{\mu\nu}$ is the trace of the extrinsic curvature. Beyond, we have that $\bar{\gamma}_{\mu\nu}$ and $\Sigma^{(\bar{\gamma})}$ are the induced metric and the surface tension on the boundary $r\to\infty$ \cite{Santos:2021orr}. Thus, we have the Euclidean action 
\begin{eqnarray}\label{Euclidean_action}
&&I_{E}=-\frac{V_{2} L^{2}r^{z+1}_{+}}{8G}\,
\Sigma^{\kappa}_{z}(\xi)\,, 
\end{eqnarray}
where the function $\Sigma^{\kappa}_{z}(\xi)$ is defined in terms of the parameters $\kappa$, $z$, and  $\xi$ as 
\begin{eqnarray}\label{Sigma_kappa_z_xi}
&&\Sigma^{\kappa}_{z}(\xi)=\frac{3}{5}+\frac{2}{5}\, \frac{z^{2}+2}{z+2}-\xi\,\frac{z+1}{z+2}\; _2F_1\left(1,\frac{z+2}{\lambda };\frac{z+2}{\lambda }+1;-\kappa^2\right)\,.
\label{K1}
\end{eqnarray}
\begin{figure}[!ht]
\begin{center}
\includegraphics[scale=0.8]{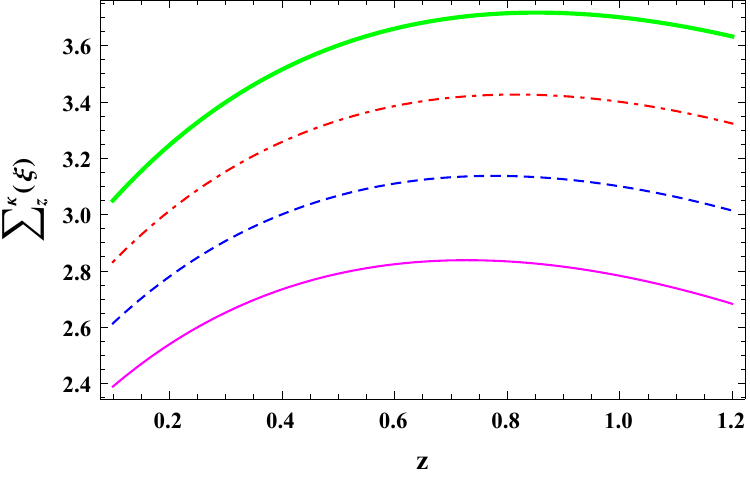}
\caption{This figure shows the behavior of the equation (\ref{K1}) considering the values $\Lambda=-1$,  $\alpha=-0.1$, $\gamma=-0.2$, which implies $\xi=10$ and $\lambda=3(1-z)$, with the choices $\kappa=0.1$ ({\sl  solid pink}), $\kappa=0.01$ ({\sl dot dashed blue}), $\kappa=0.001$ ({\sl dashed red})  and $\kappa=0.0001$ ({\sl thick green}).}\label{k1}
\label{planohwkhz1}
\end{center}
\end{figure} 
\noindent 
The profile of this function is shown in  Fig.~\ref{k1}, which can change the behavior of the relevant thermodynamic quantities, i.e., provide a change from a large to a small black hole. Thus, the mass of the black hole, which can be found through the equation $dM=TdS$ and following the procedures of \cite{Bravo-Gaete:2022lno,Hu:2019lcy,Santos:2022fbq,Dimopoulos:2001hw}, reads 
\begin{eqnarray}
M_{BH}=\frac{1}{4\pi}M^{2}_{p}\, \Sigma^{\kappa}_{z}(\xi)\, {(1+z)\, r^{2z+1}_{+}}\,, 
\label{mass}
\end{eqnarray}
where $M_{p}=G^{-1/2}$ is the  Planck mass in units in which $\hbar=c=1$, compatible with the area law. Note that when $z\to 1$ and $\xi\to 0$, then $\Sigma^{\kappa}_{z}(\xi)\to 1$ and  we reproduce the well known result of Hawking and Page \cite{Hawking:1982dh}. 
If we rewrite the black hole (brane) mass as a function of the temperature, Eq. \eqref{T}, we find:
\begin{eqnarray}
M_{BH}=\frac{1}{4\pi}M^{2}_{p}\, \Sigma^{\kappa}_{z}(\xi)\, (1+z)\,\,\left(\frac{4\pi \, T}{1+2z}\right)^{2+\frac{1}{z}}. 
\label{massT}
\end{eqnarray}
The ratio $M_{BH}/M^{2}_{p}$ against the temperature $T$ is shown in Fig. \ref{mass1}.

\begin{figure}[!ht]
\begin{center}
\includegraphics[scale=0.8]{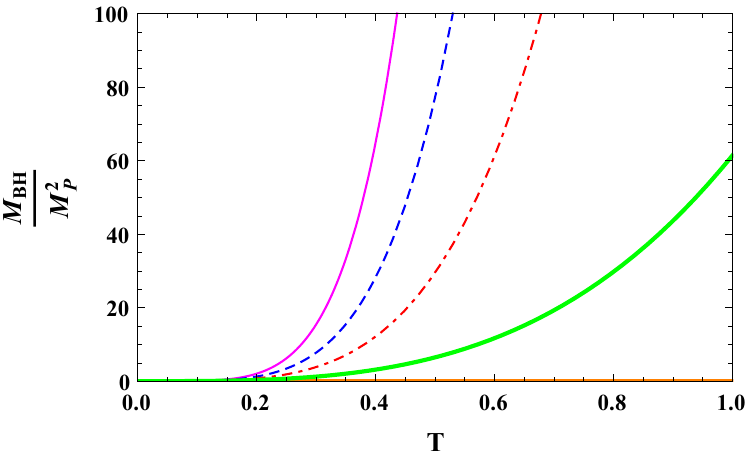}
\caption{This figure shows the behavior of the equations (\ref{mass}) considering the values $\Lambda=-1$,  $\alpha=-0.1$, $\gamma=-0.2$, which implies $\xi=10$ and $\lambda=3(1-z)$, and $\kappa=0.01$ with the choices $z=1/3$ ({\sl solid pink}), $z=2/5$ ({\sl dashed blue}), $z=1/2$ ({\sl dot dashed red}), $z=3/5$ ({\sl solid red})  and $z=4/5$ ({\sl thick green}).}\label{mass1}
\label{planohwkhz2}
\end{center}
\end{figure}

 Let us work with the canonical ensemble where the free energy is given by $F=-T\ln(Z)$, where $Z$ is the partition function defined by $Z=e^{-I_{E}}$ and $I_{E}$  is the Euclidean action, so that $F=TI_{E}$. Thus, we can write the following thermodynamic relations:
\begin{eqnarray}
\label{lnZ}
\ln(Z)&=& \sigma\Sigma^{\kappa}_{z}(\xi)\left(\frac{4\pi}{1+\frac{1}{z}}\right)^{1+\frac{1}{z}}T^{1+\frac{1}{z}}\,\\
F&=&-\sigma\Sigma^{\kappa}_{z}(\xi)\left(\frac{4\pi}{1+2z}\right)^{1+\frac{1}{z}}T^{2+\frac{1}{z}}\,, \label{15}\label{F}
\end{eqnarray}
where $\sigma\equiv{V_ {2}L^{2}}/{8G}$. Now, we can construct the thermal states in the anti-de Sitter space,  identifying the coordinate of the imaginary time $\tau$ with the period $\beta=1/T$. Thus, we can find the expected value of energy, entropy, and specific heat as
\begin{eqnarray}
\langle E\rangle&=&-\frac{\partial\ln(Z)}{\partial\beta}=\sigma\Sigma^{\kappa}_{z}(\xi)(1+1/z)\left(\frac{4\pi}{1+2z}\right)^{1+\frac{1}{z}}T^{2+\frac{1}{z}},\label{16}\label{<E>}\\
S&=&-\frac{\partial F}{\partial T}=\sigma\Sigma^{\kappa}_{z}(\xi)(2+1/z)\left(\frac{4\pi}{1+2z}\right)^{1+\frac{1}{z}}T^{1+\frac{1}{z}},\label{17}\label{S}\\
C_{V}&=&T\frac{\partial S}{\partial T}=\sigma\Sigma^{\kappa}_{z}(\xi)(2+1/z)(1+1/z)\left(\frac{4\pi}{1+2z}\right)^{1+\frac{1}{z}}T^{1+\frac{1}{z}}.\label{18}\label{CV}
\end{eqnarray}
The free energy of the black hole, its entropy, and internal energy satisfy the thermodynamic relation $F=\langle E\rangle-TS$. Considering the value of critical exponent as $z=1$, this gives us $\langle E\rangle\sim T^{3}$ that is similar to the case of energy of the thermal radiation in \cite{Hawking:1982dh}. Actually, for $z=1$ and $z=2$,  we recover the results of \cite{Hawking:1982dh} and \cite{Dimopoulos:2001hw}, respectively. 

\begin{figure}[!ht]
\begin{center}
\includegraphics[scale=0.45]{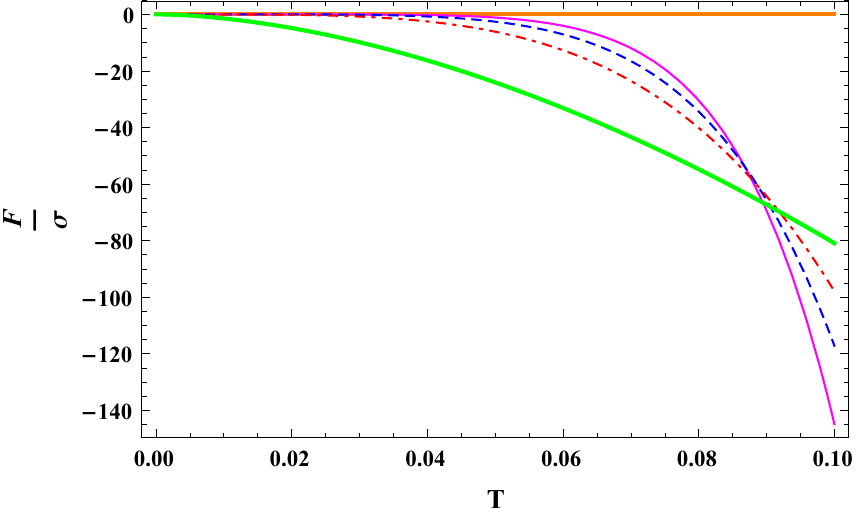}
\includegraphics[scale=0.55]{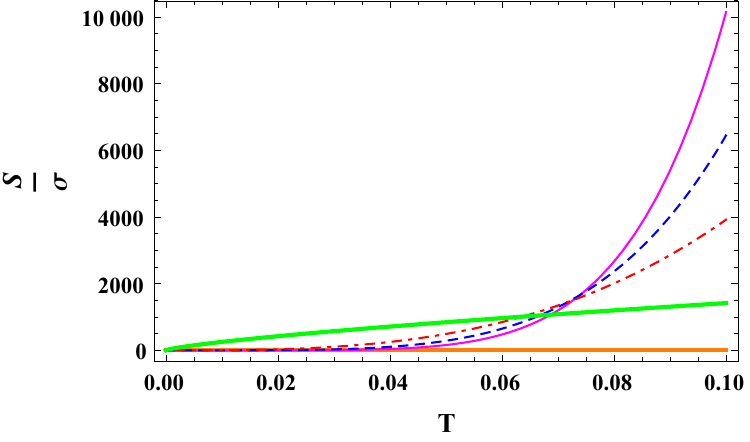}
\includegraphics[scale=0.55]{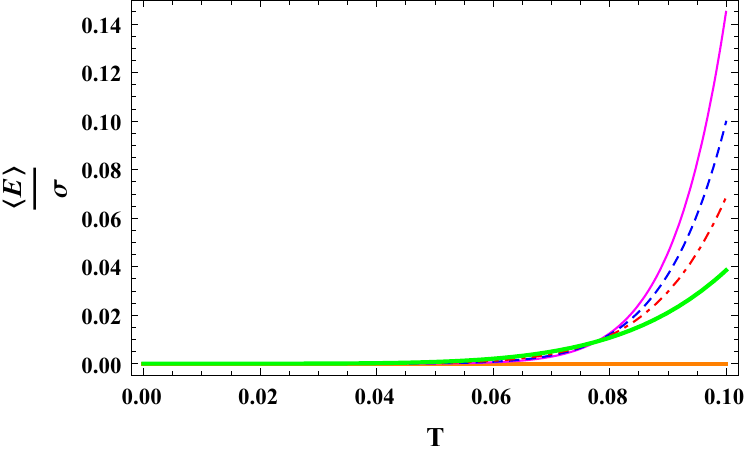}
\includegraphics[scale=0.55]{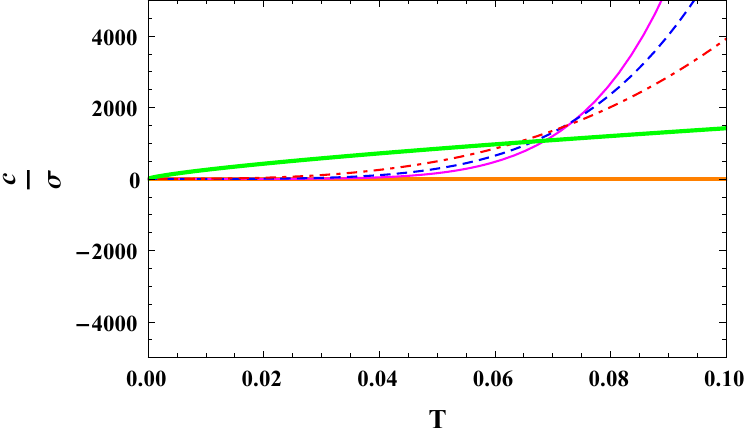}
\caption{This figure shows the behavior of the free energy $F$,  the expected value of energy $\langle E\rangle$, the entropy $S$, and the specific heat $C_V$, Eqs. (\ref{15})-(\ref{18}), respectively,    considering the values  $\Lambda=-1$,  $\alpha=-0.1$, $\gamma=-0.2$, which implies $\xi=10$ and $\lambda=3(1-z)$, and $\kappa=0.01$ with $z=1/3$ ({\sl solid pink}), $z=2/5$ ({\sl  dashed blue}), $z=1/2$ ({\sl dot dashed red}) and $z=4/5$ ({\sl thick green}).}\label{fig3}
\label{planohwkhz3}
\end{center}
\end{figure}

\begin{figure}[!ht]
\begin{center}
\includegraphics[scale=0.45]{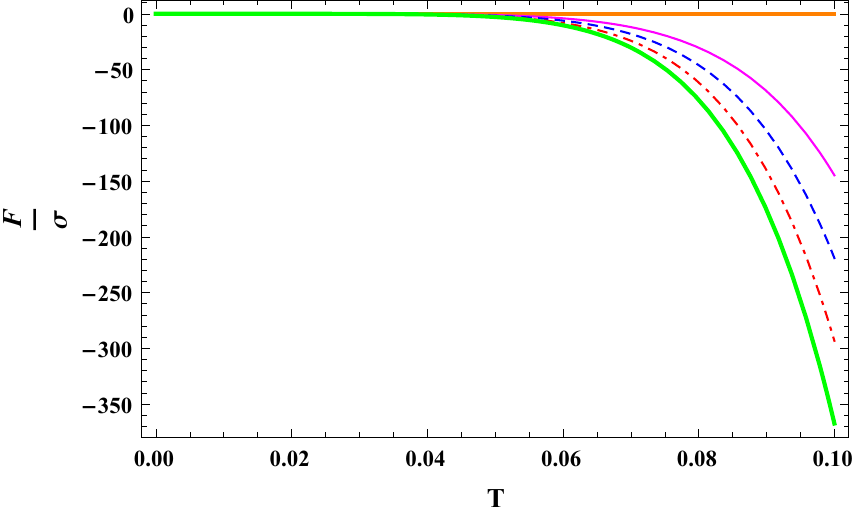}
\includegraphics[scale=0.55]{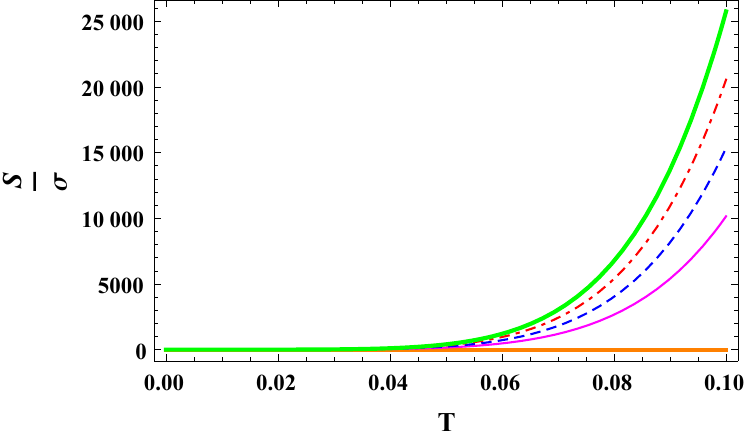}
\includegraphics[scale=0.55]{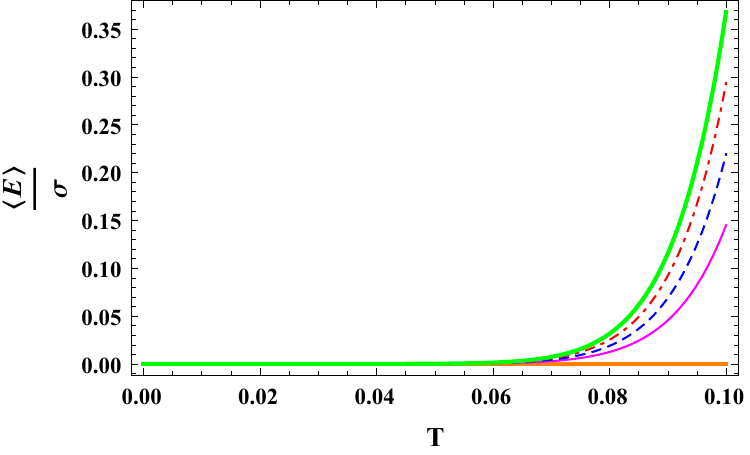}
\includegraphics[scale=0.55]{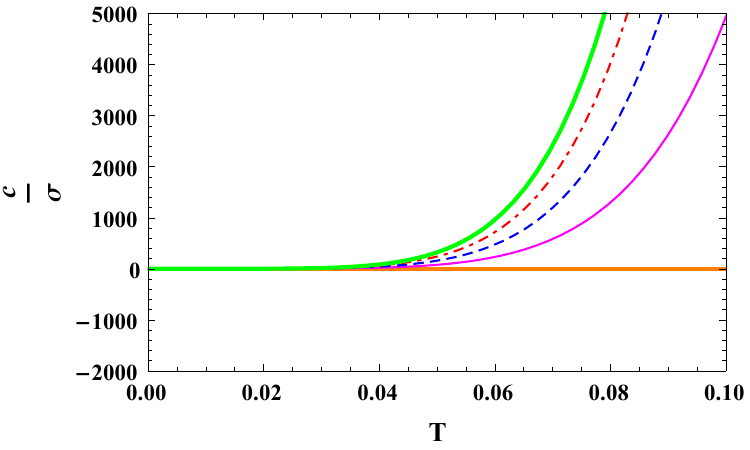}
\caption{This figure shows the behavior of the free energy $F$,  the expected value of energy $\langle E\rangle$, the entropy $S$, and the specific heat $C_V$, Eqs. (\ref{15})-(\ref{18}), respectively,  considering the values  $\Lambda=-1$,  $\alpha=-0.1$, $\kappa=0.01$, and $z=1/3$ with $\gamma=-0.2$ ({\sl solid pink}), $\gamma=-0.3$ ({\sl dot dashed blue}), $\gamma=-0.4$ ({\sl dot dashed red}), and $\gamma=-0.5$ ({\sl thick green}).}\label{fig4}
\label{planohwkhz4}
\end{center}
\end{figure}

In Figures \ref{fig3} and \ref{fig4}, we plot the free energy $F$, Eq. (\ref{15}), the expected value of energy $\langle E\rangle$, Eq. (\ref{16}), the entropy $S$, Eq. (\ref{17}), and the specific heat $C_V$, Eq. (\ref{18}) of the black branes in asymptotically Lifshitz spacetimes in $\kappa$-Horndeski gravity as functions of the temperature~$T$ for various values of the parameters $z$ and $\gamma$, respectively. 
Note that the positivity of the expected value of the energy $\langle E\rangle$ and the specific heat $C_V$ are in accordance with  \cite{Hawking:1982dh} indicating a large black hole (brane) mass, which is at least locally stable. 
In addition, the negativity of the free energy $F$ indicates that the system has global stability \cite{Rostami:2019xrx} also with a large black hole (brane) mass. 

Beyond these thermodynamics quantities, we can also extract the speed of sound and trace the energy-momentum tensor following  \cite{Ballon-Bayona:2020xls,Chen:2020ath}. Thus, we have the speed of sound as:
\begin{eqnarray}
c^{2}_{s}=\frac{\partial p}{\partial\epsilon}=\frac{\partial p/\partial r_{+}}{\partial\epsilon/\partial r_{+}}=\frac{S}{C_{V}}=\frac{z}{1+z},\label{sound}
\end{eqnarray}
where for $z=1$, we have the value to the speed of sound given by $c^{2}_{s}=1/2$, which is associated with a conformal system in \cite{Ballon-Bayona:2020xls,Chen:2020ath}.

The trace of the energy-momentum tensor reads:
\begin{eqnarray}
\langle T^a_{\ \ a}\rangle &=& \epsilon - 3p = 4F + TS,\cr
&=& -3\, \sigma\, \Sigma^{\kappa}_{z}(\xi)\left(\frac{4\pi}{1+2z}\right)^{1+\frac{1}{z}}T^{2+\frac{1}{z}}.\label{sym}
\end{eqnarray}
Therefore, the scaled trace of the energy momentum tensor $\langle\mathfrak{T}^a_{\ \ a}\rangle\equiv\langle T^a_{\ \ a}\rangle/T^4$ behaves as a function of the temperature as $T^{-2+1/z}$. Note that this scaled trace should vanish for a conformal theory. 
If $T^{-2+1/z}$ grows,  we have a broken conformal symmetry in asymptotically Lifshitz spacetimes in $\kappa$-Horndeski gravity. On the other hand, if we consider in the equation (\ref{sym}) in the particular case of the value $z=1$, this implies $\langle\mathfrak{T}^a_{\ \ a}\rangle\sim1/T$ and for the high-temperature regime, despite the influence of $\kappa$-Horndeski gravity, $\langle\mathfrak{T}^a_{\ \ a}\rangle\to 0$, which means the restoration of the conformal symmetry for this case. Other values of critical exponent also provide the conformal regime; for example large values of $z$ like $z=50$ are expected for thermal states in Lifshitz harmonic models \cite{MohammadiMozaffar:2017chk}; with respect to this prescription, if consider $z\to \infty$ the scaled trace $\langle\mathfrak{T}^a_{\ \ a}\rangle$ and the trace itself $\langle T^a_{\ \ a}\rangle$ vanish identically independent of the temperature and the conformal symmetry is exact in this case.

Before we finish this section, it is interesting to note that ratios of the free energy, Eq. \eqref{15}, and the expected value of energy $\langle E\rangle$, Eq. (\ref{16}), with respect to the black (hole) brane mass $M_{BH}$, Eq. \eqref{massT} do not dependent on the temperature or the parameters $\kappa$ and $\xi$:   
\begin{eqnarray}
\label{FM}
\frac{F}{M_{BH}}&=&-\frac{\sigma}{M_p^2}\frac{1+2z}{1+z}\,, \\ 
\frac{\langle E\rangle}{M_{BH}}&=&\frac{\sigma}{M_p^2} \frac{(1+1/z)(1+2z)}{1+z}\,. 
\label{EM}
\end{eqnarray}
Then, we see that the ratio of the free energy $F$ and the expected value of energy $\langle E\rangle$, is simply given by
\begin{eqnarray}
\frac{F}{\langle E\rangle}&=&-
 (1+1/z) \, \,. 
\end{eqnarray}

 Further, the ratios of the entropy $S$, Eq. \eqref{17}, and the specific heat $C_V$, Eq. (\ref{18}) of the black branes in asymptotically Lifshitz spacetimes in $\kappa$-Horndeski gravity with respect to the black brane (hole) mass have a simple dependency on the inverse of the temperature $T$ and are independent of the parameters $\kappa$ and $\xi$:
\begin{eqnarray}
\label{SM}
\frac{S}{M_{BH}}&=&\frac{\sigma}{M_p^2} \frac{(2+1/z)(1+2z)}{1+z}\frac{1}{T}, \\
\frac{C_V}{M_{BH}}&=&\frac{\sigma}{M_p^2} \frac{(2+1/z)(1+1/z)(1+2z)}{1+z}\frac{1}{T}\,,
\label{CVM}
\end{eqnarray}
 which leads to 
\begin{eqnarray}
C_V&=&  (1+1/z)\, S \,,
\end{eqnarray}
in agreement with the square of the speed of sound, Eq. \eqref{sound}. 

These results are also consistent with the ratios $F/\langle E\rangle$ and $S/C_V$ that can be inferred, for instance, from Figs. \ref{fig3} and \ref{fig4}. 

As a final check, we can calculate the quantity 
$\langle E\rangle -TS $, from the results, Eqs. \eqref{EM} and \eqref{SM}, and find
\begin{eqnarray}
\langle E\rangle -TS 
&=&-\frac{\sigma\, M_{BH}}{M_p^2}\frac{1+2z}{1+z}\,, 
\label{Fcheck}
\end{eqnarray}
which is exactly the free energy $F$ given by Eq. \eqref{FM}, as expected.



\section{Conclusions}\label{v4}

 We have obtained and studied the thermodynamics of black brane solutions in asymptotically Lifshitz spacetimes in  $\kappa$-deformed Horndeski gravity in four dimensions. The main property of these solutions is that they can assume arbitrary values of the critical exponents $z$ associated with Lorentz symmetry violation. Usually, when asymptotically Lifshitz spacetimes are considered in Horndeski gravity, the critical exponents typically acquire some specific fixed values for which some solution can be found. What we have shown here is that the inclusion of the $\kappa$-deformation in Horndesky theory releases these constraints on the critical exponents allowing them to assume arbitrary values and then finding solutions in this general case. This is a consequence of the auxiliary equation (\ref{9.1}) that comes from the $\kappa$-algebra leading to the solution of the equations of motion for any value of the critical exponent. 
 
 Then, we have studied the thermodynamics of these black brane solutions. From the Euclidean action, Eq. \eqref{EA}, we have found the partition function, Eq. \eqref{lnZ}, from which we obtain, for instance, the free energy, the expected value of the energy,  the entropy,  and the specific heat, Eqs. \eqref{F}-\eqref{CV}, respectively, in terms of the parameters $\Lambda$,  $\alpha$, $\gamma$, $\kappa$, $z$, and the temperature $T$. We have plotted these thermodynamical functions  in Figs. \ref{fig3} and \ref{fig4}, for some values of those parameters characterizing the asymptotically Lifshitz spacetimes in  $\kappa$-deformed Horndeski gravity in four dimensions.

Furthermore, we have obtained the ratio of these thermodynamical functions with respect to the black brane mass, Eq. \eqref{mass},  once this mass is itself a thermodynimcal function of the temperature $T$ and the parameters $\Lambda$,  $\alpha$, $\gamma$, $\kappa$, and $z$. These ratios provide simpler expressions for the the free energy, the expected value of the energy,  the entropy,  and the specific heat, given by Eqs. in Eqs. \eqref{FM}, \eqref{EM}, \eqref{SM}, \eqref{CVM}, respectively. From these expressions, we could check that they are consistent with the speed of sound, Eq. \eqref{sound} and that the thermodynamical relation 
 $F=\langle E\rangle-TS$, Eq. \eqref{Fcheck}, holds too, as expected. 

Finally, the the positivity of the expected value of the energy $\langle E\rangle$ and the specific heat $C_V$  indicate at least local stability and the negativity of the free energy $F$ shows that the system has global stability  both compatible with a large black hole (brane) mass \cite{Hawking:1982dh, Rostami:2019xrx}.

We hope that this study might help finding different applications for Horndeski gravity in asymptotically Lifshitz spacetimes.

\section*{Acknowledgments}

 We would like to thank Mokhtar Hassine for the fruitful discussions. This work was supported by Coordenação de Aperfeiçoamento de Pessoal de Nível Superior (CAPES) under finance code 001.  H.B.-F. is partially supported by Conselho Nacional de Desenvolvimento Cient\'{\i}fico e Tecnol\'{o}gico (CNPq) under grant $\#$ 311079/2019-9.



\begin{thebibliography}{99}

\bibitem{Horndeski:1974wa} 
G.~W.~Horndeski,
{\it Second-order scalar-tensor field equations in a four-dimensional space},
Int.\ J.\ Theor.\ Phys.\  {\bf 10}, 363 (1974).
doi:10.1007/BF01807638. 


\bibitem{Charmousis:2011bf} 
  C.~Charmousis, E.~J.~Copeland, A.~Padilla and P.~M.~Saffin,
  {\it General second order scalar-tensor theory, self tuning, and the Fab Four},
  Phys.\ Rev.\ Lett.\  {\bf 108}, 051101 (2012), [arXiv:1106.2000 [hep-th]].
	
\bibitem{Charmousis:2011ea} 
  C.~Charmousis, E.~J.~Copeland, A.~Padilla and P.~M.~Saffin,
  {\it Self-tuning and the derivation of a class of scalar-tensor theories},
  Phys.\ Rev.\ D {\bf 85}, 104040 (2012), [arXiv:1112.4866 [hep-th]].


\bibitem{Bruneton:2012zk} 
  J.~P.~Bruneton, M.~Rinaldi, A.~Kanfon, A.~Hees, S.~Schlogel and A.~Fuzfa,
  {\it Fab Four: When John and George play gravitation and cosmology},
  Adv.\ Astron.\  {\bf 2012}, 430694 (2012),
  [arXiv:1203.4446 [gr-qc]].
	

\bibitem{Heisenberg:2018vsk} 
  L.~Heisenberg,
  {\it A systematic approach to generalisations of General Relativity and their cosmological implications},
  Phys.\ Rept.\  {\bf 796}, 1 (2019),
  [arXiv:1807.01725 [gr-qc]].
 

\bibitem{Kobayashi:2019hrl}
T.~Kobayashi,
``Horndeski theory and beyond: a review,''
Rept. Prog. Phys. \textbf{82}, no.8, 086901 (2019)
doi:10.1088/1361-6633/ab2429
[arXiv:1901.07183 [gr-qc]].


\bibitem{LIGOScientific:2017ync}
B.~P.~Abbott \textit{et al.},
``Multi-messenger Observations of a Binary Neutron Star Merger,''
Astrophys. J. Lett. \textbf{848}, no.2, L12 (2017)
[arXiv:1710.05833 [astro-ph.HE]].


\bibitem{Oikonomou:2020sij}
V.~K.~Oikonomou and F.~P.~Fronimos,
``Reviving non-minimal Horndeski-like theories after GW170817: kinetic coupling corrected Einstein\textendash{}Gauss\textendash{}Bonnet inflation,'' 
Class. Quant. Grav. \textbf{38}, no.3, 035013 (2021)
doi:10.1088/1361-6382/abce47
[arXiv:2006.05512 [gr-qc]].




\bibitem{Rinaldi:2012vy} 
  M.~Rinaldi,
  {\it Black holes with non-minimal derivative coupling},
  Phys.\ Rev.\ D {\bf 86}, 084048 (2012),
  [arXiv:1208.0103 [gr-qc]].

  \bibitem{Gleyzes:2013ooa} 
  J.~Gleyzes, D.~Langlois, F.~Piazza and F.~Vernizzi,
  {\it Essential Building Blocks of Dark Energy},
  JCAP {\bf 1308}, 025 (2013),
  [arXiv:1304.4840 [hep-th]].
	
\bibitem{Zumalacarregui:2013pma} 
  M.~Zumalacárregui and J.~García-Bellido,
  {\it Transforming gravity: from derivative couplings to matter to second-order scalar-tensor theories beyond the Horndeski Lagrangian},
  Phys.\ Rev.\ D {\bf 89}, 064046 (2014),
  [arXiv:1308.4685 [gr-qc]].

\bibitem{Cisterna:2016vdx} 
  A.~Cisterna, T.~Delsate, L.~Ducobu and M.~Rinaldi,
  {\it Slowly rotating neutron stars in the nonminimal derivative coupling sector of Horndeski gravity},
  Phys.\ Rev.\ D {\bf 93}, no. 8, 084046 (2016),
  [arXiv:1602.06939 [gr-qc]].

\bibitem{Starobinsky:2016kua} 
  A.~A.~Starobinsky, S.~V.~Sushkov and M.~S.~Volkov,
  {\it The screening Horndeski cosmologies},
  JCAP {\bf 1606}, no. 06, 007 (2016),
  [arXiv:1604.06085 [hep-th]].


  

\bibitem{Rinaldi:2016oqp} 
  M.~Rinaldi,
  {\it Mimicking dark matter in Horndeski gravity},
  Phys.\ Dark Univ.\  {\bf 16}, 14 (2017),
  [arXiv:1608.03839 [gr-qc]].
	
\bibitem{Brito:2018pwe}
F.~A.~Brito and F.~F.~Santos,
``Braneworlds in Horndeski gravity,''
Eur. Phys. J. Plus \textbf{137}, no.9, 1051 (2022)
doi:10.1140/epjp/s13360-022-03270-w
[arXiv:1810.08196 [hep-th]].
		




\bibitem{Maldacena:1997re} 
  J.~M.~Maldacena,
  {\it The Large N limit of superconformal field theories and supergravity},
  Int.\ J.\ Theor.\ Phys.\  {\bf 38}, 1113 (1999)
  [Adv.\ Theor.\ Math.\ Phys.\  {\bf 2}, 231 (1998)],
  [hep-th/9711200].

\bibitem{Aharony:1999ti}
O.~Aharony, S.~S.~Gubser, J.~M.~Maldacena, H.~Ooguri and Y.~Oz,
``Large N field theories, string theory and gravity,''
Phys. Rept. \textbf{323}, 183-386 (2000)
doi:10.1016/S0370-1573(99)00083-6
[arXiv:hep-th/9905111 [hep-th]].

\bibitem{Petersen:1999zh}
J.~L.~Petersen,
``Introduction to the Maldacena conjecture on AdS / CFT,''
Int. J. Mod. Phys. A \textbf{14}, 3597-3672 (1999)
doi:10.1142/S0217751X99001676
[arXiv:hep-th/9902131 [hep-th]].

\bibitem{Ramallo:2013bua}
A.~V.~Ramallo,
``Introduction to the AdS/CFT correspondence,''
Springer Proc. Phys. \textbf{161}, 411-474 (2015)
doi:10.1007/978-3-319-12238-0\_10
[arXiv:1310.4319 [hep-th]].


\bibitem{Policastro:2001yc}
G.~Policastro, D.~T.~Son and A.~O.~Starinets,
``The Shear viscosity of strongly coupled N=4 supersymmetric Yang-Mills plasma,'' 
Phys. Rev. Lett. \textbf{87}, 081601 (2001)
doi:10.1103/PhysRevLett.87.081601
[arXiv:hep-th/0104066 [hep-th]].


\bibitem{Kovtun:2003wp} 
  P.~Kovtun, D.~T.~Son and A.~O.~Starinets,
  {\it Holography and hydrodynamics: Diffusion on stretched horizons},
  JHEP {\bf 0310}, 064 (2003),
  [hep-th/0309213].

  
\bibitem{Kovtun:2004de}
P.~Kovtun, D.~T.~Son and A.~O.~Starinets,
``Viscosity in strongly interacting quantum field theories from black hole physics,'' 
Phys. Rev. Lett. \textbf{94}, 111601 (2005)
doi:10.1103/PhysRevLett.94.111601
[arXiv:hep-th/0405231 [hep-th]].

\bibitem{Hartnoll:2009sz} 
  S.~A.~Hartnoll,
  {\it Lectures on holographic methods for condensed matter physics},
  Class.\ Quant.\ Grav.\  {\bf 26}, 224002 (2009),
  [arXiv:0903.3246 [hep-th]].
  
\bibitem{Sachdev:2011wg} 
  S.~Sachdev,
  {\it What can gauge-gravity duality teach us about condensed matter physics?},
  Ann.\ Rev.\ Condensed Matter Phys.\  {\bf 3}, 9 (2012),
[arXiv:1108.1197 [cond-mat.str-el]].

\bibitem{Anabalon:2013oea} 
A.~Anabalon, A.~Cisterna and J.~Oliva,
{\it Asymptotically locally AdS and flat black holes in Horndeski theory},
Phys.\ Rev.\ D {\bf 89}, 084050 (2014),
[arXiv:1312.3597 [gr-qc]].


\bibitem{Cisterna:2014nua} 
A.~Cisterna and C.~Erices,
{\it Asymptotically locally AdS and flat black holes in the presence of an electric field in the Horndeski scenario},
Phys.\ Rev.\ D {\bf 89}, 084038 (2014),
[arXiv:1401.4479 [gr-qc]].

\bibitem{Cisterna:2017jmv} 
  A.~Cisterna, M.~Hassaine, J.~Oliva and M.~Rinaldi,
  {\it Axionic black branes in the k-essence sector of the Horndeski model},
  Phys.\ Rev.\ D {\bf 96}, no. 12, 124033 (2017),
  [arXiv:1708.07194 [hep-th]].

\bibitem{Feng:2015oea} 
X.~H.~Feng, H.~S.~Liu, H.~Lü and C.~N.~Pope,
{\it Black Hole Entropy and Viscosity Bound in Horndeski Gravity},
JHEP {\bf 1511}, 176 (2015),
[arXiv:1509.07142 [hep-th]].




  
\bibitem{Horava:2008ih}
P.~Horava,
``Membranes at Quantum Criticality,''
JHEP \textbf{03}, 020 (2009)
doi:10.1088/1126-6708/2009/03/020
[arXiv:0812.4287 [hep-th]].

\bibitem{Horava:2009if}
P.~Horava,
``Spectral Dimension of the Universe in Quantum Gravity at a Lifshitz Point,''
Phys. Rev. Lett. \textbf{102}, 161301 (2009)
doi:10.1103/PhysRevLett.102.161301
[arXiv:0902.3657 [hep-th]].

\bibitem{Horava:2010zj}
P.~Horava and C.~M.~Melby-Thompson,
``General Covariance in Quantum Gravity at a Lifshitz Point,''
Phys. Rev. D \textbf{82}, 064027 (2010)
doi:10.1103/PhysRevD.82.064027
[arXiv:1007.2410 [hep-th]].

\bibitem{Barausse:2011pu}
E.~Barausse, T.~Jacobson and T.~P.~Sotiriou,
``Black holes in Einstein-aether and Horava-Lifshitz gravity,'' 
Phys. Rev. D \textbf{83}, 124043 (2011)
doi:10.1103/PhysRevD.83.124043
[arXiv:1104.2889 [gr-qc]].

\bibitem{Wang:2017brl}
A.~Wang,
``Ho\v{r}ava gravity at a Lifshitz point: A progress report,'' 
Int. J. Mod. Phys. D \textbf{26}, no.07, 1730014 (2017)
doi:10.1142/S0218271817300142
[arXiv:1701.06087 [gr-qc]].


\bibitem{Clifton:2011jh}
T.~Clifton, P.~G.~Ferreira, A.~Padilla and C.~Skordis,
``Modified Gravity and Cosmology,''
Phys. Rept. \textbf{513}, 1-189 (2012)
doi:10.1016/j.physrep.2012.01.001
[arXiv:1106.2476 [astro-ph.CO]].






\bibitem{Danielsson:2009gi}
U.~H.~Danielsson and L.~Thorlacius,
``Black holes in asymptotically Lifshitz spacetime,''
JHEP \textbf{03}, 070 (2009)
doi:10.1088/1126-6708/2009/03/070
[arXiv:0812.5088 [hep-th]].

\bibitem{Bertoldi:2009vn}
G.~Bertoldi, B.~A.~Burrington and A.~Peet,
``Black Holes in asymptotically Lifshitz spacetimes with arbitrary critical exponent,''
Phys. Rev. D \textbf{80}, 126003 (2009)
doi:10.1103/PhysRevD.80.126003
[arXiv:0905.3183 [hep-th]].



\bibitem{Bravo-Gaete:2013dca} 
M.~Bravo-Gaete and M.~Hassaine,
{\it Lifshitz black holes with a time-dependent scalar field in a Horndeski theory},
Phys.\ Rev.\ D {\bf 89}, 104028 (2014),
[arXiv:1312.7736 [hep-th]].


\bibitem{Santos:2020egn}
F.~F.~Santos,
{\it Aplica\c{c}\~oes do Setor John da Gravidade de Horndeski nos Cen\'arios de Brana Negra e Rela\c{c}\~ao de viscosidade/entropia, Mundo Brana e Cosmologia (In Portuguese)},
[arXiv:2006.06550 [hep-th]].
  





\bibitem{Brito:2019ose}
F.~A.~Brito and F.~F.~Santos,
{\it Black brane in asymptotically Lifshitz spacetime and viscosity/entropy ratios in Horndeski gravity}, 
EPL \textbf{129}, no.5, 50003 (2020)
doi:10.1209/0295-5075/129/50003
[arXiv:1901.06770 [hep-th]].


\bibitem{Santos:2022fbq}
F.~F.~Santos, B.~G.~da Costa and I.~S.~Gomez,
{\it Studies of transport coefficients in charged AdS$_{4}$ black holes on $\kappa$-deformed space},
doi:10.22128/JHAP.2022.608.1035
[arXiv:2211.13783 [hep-th]].










\bibitem{daCosta:2020mbf}
B.~G.~da Costa, I.~S.~Gomez and M.~Portesi,
{\it $\kappa$-Deformed quantum and classical mechanics for a system with position-dependent effective mass},
J. Math. Phys. \textbf{61}, no.8, 082105 (2020),
[arXiv:2007.11184 [quant-ph]].



\bibitem{Kaniadakis}
G.~Kaniadakis
{\it Non-linear kinetics underlying generalized statistics},
Physica A \textbf{296} (2001), 405-425,
[arxiv:cond-mat/0103467[cond-mat.stat-mech]]


\bibitem{Deffayet:2011gz} 
  C.~Deffayet, X.~Gao, D.~A.~Steer and G.~Zahariade,
  {\it From k-essence to generalised Galileons},
  Phys.\ Rev.\ D {\bf 84}, 064039 (2011),
  [arXiv:1103.3260 [hep-th]].


\bibitem{Borissova:2022mgd}
J.~N.~Borissova and A.~Platania,
{\it Formation and evaporation of quantum black holes from the decoupling mechanism in quantum gravity},  
JHEP \textbf{03}, 046 (2023)
doi:10.1007/JHEP03(2023)046
[arXiv:2210.01138 [gr-qc]].

\bibitem{Hui:2012qt}
L.~Hui and A.~Nicolis,
``No-Hair Theorem for the Galileon,'' 
Phys. Rev. Lett. \textbf{110}, 241104 (2013)
doi:10.1103/PhysRevLett.110.241104
[arXiv:1202.1296 [hep-th]].


\bibitem{Santos:2021orr}
F.~F.~Santos, E.~F.~Capossoli and H.~Boschi-Filho,
{\it AdS/BCFT correspondence and BTZ black hole thermodynamics within Horndeski gravity},
Phys. Rev. D \textbf{104}, no.6, 066014 (2021),
[arXiv:2105.03802 [hep-th]].


\bibitem{Jiang:2017imk} 
 W.~J.~Jiang, H.~S.~Liu, H.~Lu and C.~N.~Pope, 
 {\it DC Conductivities with Momentum Dissipation in Horndeski Theories}, 
 JHEP {\bf 1707}, 084 (2017), 
 [arXiv:1703.00922 [hep-th]].


\bibitem{Bekenstein:1973ur}
J.~D.~Bekenstein, 
``Black holes and entropy,'' 
Phys. Rev. D \textbf{7}, 2333-2346 (1973)
doi:10.1103/PhysRevD.7.2333

\bibitem{Hawking:1975vcx}
S.~W.~Hawking,
``Particle Creation by Black Holes,'' 
Commun. Math. Phys. \textbf{43}, 199-220 (1975)
[erratum: Commun. Math. Phys. \textbf{46}, 206 (1976)]
doi:10.1007/BF02345020


\bibitem{Gibbons:1976ue}
G.~W.~Gibbons and S.~W.~Hawking,
``Action Integrals and Partition Functions in Quantum Gravity,'' 
Phys. Rev. D \textbf{15}, 2752-2756 (1977)
doi:10.1103/PhysRevD.15.2752


\bibitem{Hu:2019lcy}
S.~Q.~Hu, Y.~C.~Ong and D.~N.~Page,
{\it No evidence for violation of the second law in extended black hole thermodynamics},
Phys. Rev. D \textbf{100}, no.10, 104022 (2019)
doi:10.1103/PhysRevD.100.104022
[arXiv:1906.05870 [gr-qc]].


\bibitem{Bravo-Gaete:2022lno}
M.~Bravo-Gaete, F.~F.~Santos and H.~Boschi-Filho,
{\it Shear viscosity from black holes in generalized scalar-tensor theories in arbitrary dimensions},''
Phys. Rev. D \textbf{106}, no.6, 066010 (2022)
doi:10.1103/PhysRevD.106.066010
[arXiv:2201.07961 [hep-th]].


\bibitem{Dimopoulos:2001hw} 
 S.~Dimopoulos and G.~L.~Landsberg,
{\it Black holes at the LHC},
Phys.\ Rev.\ Lett.\  {\bf 87}, 161602 (2001),
[hep-ph/0106295].

\bibitem{Hawking:1982dh} 
 S.~W.~Hawking and D.~N.~Page,
{\it Thermodynamics of Black Holes in anti-De Sitter Space},
 Commun.\ Math.\ Phys.\  {\bf 87}, 577 (1983).


\bibitem{Rostami:2019xrx} 
  M.~Rostami, J.~Sadeghi, S.~Miraboutalebi, B.~Pourhassan and A.~A.~Masoudi,
  {\it Phase transition of modified Horndeski gravity with new method},
  arXiv:1909.01799 [gr-qc].


\bibitem{Ballon-Bayona:2020xls} 
  A.~Ballon-Bayona, H.~Boschi-Filho, E.~F.~Capossoli and D.~M.~Rodrigues,
  {\it Criticality from Einstein-Maxwell-dilaton holography at finite temperature and density},''
  Phys.\ Rev.\ D {\bf 102}, no. 12, 126003 (2020),
  [arXiv:2006.08810 [hep-th]].
  

\bibitem{Chen:2020ath}
X.~Chen, L.~Zhang, D.~Li, D.~Hou and M.~Huang,
``Gluodynamics and deconfinement phase transition under rotation from holography,'' 
JHEP \textbf{07}, 132 (2021)
doi:10.1007/JHEP07(2021)132
[arXiv:2010.14478 [hep-ph]].

\bibitem{MohammadiMozaffar:2017chk}
M.~R.~Mohammadi Mozaffar and A.~Mollabashi,
{\it Logarithmic Negativity in Lifshitz Harmonic Models},
J. Stat. Mech. \textbf{1805}, no.5, 053113 (2018)
doi:10.1088/1742-5468/aac135
[arXiv:1712.03731 [hep-th]].



  


  









 





			
\end{thebibliography}
\end{document}